# Comparative Analysis of Dynamic Graph Techniques and Data Structure

Deepak Garg
Associate Professor
CSED, Thapar University
Patiala, India

Megha Tyagi
Student- M.E
CSED, Thapar University
Patiala, India

## ABSTRACT

Dynamically changing graphs are used in many applications of graph algorithms. The scope of these graphs are in graphics, communication networks and in VLSI designs where graphs are subjected to change, such as addition and deletion of edges and vertices. There is a rich body of the algorithms and data structures used for dynamic graphs. The paper overview the techniques and data structures used in various dynamic algorithms. The effort is tried to find out the comparison in these techniques namely the hierarchical decomposition of graphs and highlighting the ingenuity used in designing these algorithms.

## 1. INTRODUCTION TO DYNAMIC GRAPH ALGORITHM

**Dynamic Graph**

A Graph is a collection of nodes and edges which represent the network of entities and association between these nodes.

**Definition**: A dynamic graph $G=\{G_0, G_1,\ldots G_m\}$ is a sequence of graphs, where $G_s = (X_s, Y_s)$ represent the instance of graph at any time T.

- when the new links are added and deleted from the graph

According to the operation supported, the algorithm can be divided into two categories:

- A fully dynamic graph algorithm: this algorithm supports both the insertion and deletion of edges.
- Partially dynamic graph algorithms: this type of algorithm support only edge insertion or deletion but not both.

The dynamic graph algorithms are the prominent area of research for last few decades and various algorithms has been developed to maintain the graph properties like minimum spanning tree, planarity, 2-edge connectivity and bipartiteness. As dynamic graph algorithms perform better than their static counterparts, they are more difficult to design and analyse.

Dynamic graph algorithm handle the graph problems, where the graph undergoes the series of updates including the insertion of an edge and deletion of an edge and answers the various queries like Whether the graph is connected or not. The algorithm find out the solution for the various updates and queries and perform better than the static algorithm which answer to the solution computing from the Scratch. Hence the dynamic graph algorithm does not require the whole previously computed information about the graph and improves the lower bound as comparable to their static counterparts.

**Definition** : The Dynamic Algorithm compute some function X on the initial input Y and maintain the detail about X(Y) where Y is a initial input ,and work without re-evaluating X(Y) from starting as does in static algorithms.

The Dynamic graph can be visualised as a world wide web where the graph vertex represent the nodes and edges represents the links in the graph. The web graph changes dynamically as many nodes and links losses functionality as the network becomes larger.

The dynamic graph algorithm provides answer to the following operation.

- Whether the two nodes are connected or not?
- Retain the various graph properties like minimum spaning forest, vertex connectivity and bipartiteness etc.Various update take place in dynamic graph

## 2. TECHNIQUES OF DYNAMIC GRAPH ALGORITHMS

The main aspect of the paper is to analyze the various techniques which are used in dynamic graph algorithm to speed up the graph. The techniques can be applied in a wide variety of problems including finding out the minimum spanning forest, vertex and edge connectivity, planarity, and for finding out the transitive closure of the graph.

The techniques considered are:
- Clustering
- Sparsification and
- Randomization

### 2.1 Clustering

The clustering is the subdivision of graph node set into groups. This technique firstly introduced by Frederickson [1], and which partition the graph into a smaller subdivision of connected sub graph called **clusters**. The techniques use the tree data structure to store the information about the graph edges and nodes.

**Definition**: A clustering Ç (G) of a graph G=(X, Y) is a subdivision of vertices X into disarranges, nonempty subset of $\{C_1, C_2, C_3 .....C_k\}$ where $C_i £ Ç$.

The technique work as follow:

- It is based upon the decomposition of vertex set V into the sub graph called clusters and the decomposition applied recursively to the higher level. And the information about the sub graph is combined with topology tree [3].
- The clustering technique improves further in which the edges can be in multiple groups, and only one edge will be selected depending upon the topology of the spanning tree.

Dynamic clustering can be defined as:

**Definition**: A dynamic clustering Ç (G) = $\{C_1, C_2, C_3 .....C_L\}$ of a dynamic graph G, with length L, consist of a set Clustering $C_1, C_2, C_3.....C_L$ where $C_i$ is a clustering of a graph $G_i$.

Clustering when used for a single level in dynamic graph algorithms obtain the lower bound of $O(m^{2/3})$ but when the partition is applied recursively to the higher level using the two dimensional topology tree the lower time bound improved to $O(m^{1/2})$.

**Advantages**: the technique work faster for the Dynamic graph algorithm and is suitable for the deterministic algorithm. Clustering can be in cooperated with other graph technique to produce the efficient results. The algorithm has lower search space in dynamic approach and has a quick response to the clustering events.

**Drawback:** To the large extent the technique is problem dependent and can be applied as a black box for dynamic algorithms.

## 2.2 Sparsification

This technique was introduced by Epstein et al [2] and it is a general technique which can be used as a black box in designing algorithms. The technique reduces the number of edges in the graph and speedup the dynamic algorithm. Due to this technique the time bound of the algorithm improves and become analogous to the sparse graphs. The technique works on the top of the given algorithm and does not demand the structural detail of the graph.

The technique makes use of certificate to be applied on the graphs. The definition is a as follows:

### 2.2.1 Definitions

**Certificate**: For any graph property P, and graph G, a certificate for G is a graph G' such that G has property if and only if G' has the property P. [2]

**Strong Certificate**: For any graph property P, and graph G, a strong certificate for G is a graph G' on the same vertex set such that, for any H, G U H has property if and only if G' U H' has the property P. [2]

**Sparse Certificate**: A strong certificate with at most cn edges on a graph G which has n vertices for some constant c.[2]

The technique work as follow:

- The graph with E edges and V nodes partition the edge of graph G into a assembly of O (E/V) sparse sub graphs where each sub graph is a order of O (V).
- The graph is then decomposed into sparser sub graph incorporating the meaningful information for each sub graph and having the sparse certificate. Hence each node in the tree is represented by the sparse certificate.
- Now when any insertion or deletion take place for edges the O (E/V) graphs with O (V) links each would be required for updates.

Let A be the algorithm that maintain some number of properties on the dynamic graph G with time bound F (E, V), where E is a number of edges and V is a number of vertex set. So the sparsification advance the decomposition of G into smaller sub graph with O (V) edges each.

Hence the technique uses the dynamic algorithm to only some small sub graph of G, resulting into advanced time bound of F (n, O (n)).

So, the techniques improves the time bound to $O(n^{1/2})$[2] where the previously known time bound were $O(m^{1/2})$ for the update operation.

**Advantages:** The technique has the following advantages:

- The techniques applied to the wide variety of graph problems comprising vertex and edge connectivity, minimum spanning forest and bipartite graphs. As an example, for the fully dynamic minimum spanning tree problem, it reduces the update time from O (PE) to O (PV).
- The technique speed up the dynamic graph algorithm and work superior for small update sequences.
- It provides the improved space usage of O (ElogV) compared to other graph techniques.

## 2.3 Randomization

The third technique introduced by Henzinger and king [3] for dynamic graph algorithm uses the power of randomization for improving the faster update time. In this technique the graph decomposition take place with randomization.

The technique advance the lower time bound for fully dynamic graph algorithm for properties like connectivity bipartiteness and minimum spanning forest of a graph. The result of this technique achieves the faster fully and partially dynamic algorithm.

### 2.3.1 Random Sampling

The random sampling is the key idea behind this technique. When any of the edge e is removed from the tree then the edges (non tree edges) which are incident on the tree T will be randomly selected for the replacement of the deleted edge.

The main idea of the technique:

- The graph is decomposed to the O (logn) level. Where the dense part (highly connected) of the graph is connected to the lower level than those where the graph is sparse (weakly connected).
- When the tree edges are deleted at level i,
  There exit the two cases:
  **Case 1:** then the high probability edge is randomly selected to recombine two disjoint sub trees using random sampling.
  **Case 2:** when by deletion of the edge the graph become sparse, and random sampling fails then These edges are moved to level (i+1) and the same procedure is applied recursively on level (i+1).

Hence, the technique maintains the spanning forests for the graph, for each level i, whose edges are in level i and are below it. The technique implements the algorithm for various properties of Dynamic algorithms using the eulerian tour implementation of the spanning tree.

## 3. DATA STRUCTURE TOOLS FOR IMPLEMENTING GRAPH TECHNIQUES

There are many fully dynamic data structure for the dynamic graph problems. Many use the concept of partitioning of vertex into a disjoint set of paths. Some of the data structures are:
- Topology tree
- ET trees
- Dynamic Trees

### 3.1 Topology Trees

The tree represents the hierarchy of the tree T. these trees was introduced by fredricson ([1] [14]) to maintain the updates for dynamic trees. The tree uses the following terminology:
- Vertex cluster: connected sub graph of the Tree T.
- Cardinality: number of vertices in cluster.

**Definition**: The tree T defines the systematic division of the other tree, according to the topology.

For the restrained multilevel division, topology tree define the following properties:

- All the nodes at level $\ell \geq 1$ has at most two children, which shows the node clustering at level $\ell - 1$ union define the vertex cluster that node represents.

The tree uses the partition of vertices into clusters at each level where restricted partition is defined as:
- Every Cluster with outer degree three must have a cardinality of 1.else if outer degree for every cluster is less than three than cardinality must be of 2.
- No two neighboring cluster can be incorporated and still satisfying the above condition.

### 3.1.1 Insertion and deletion in tree

When deletion of the edge e take place in the tree T, then the deletion makes a tree divided into two trees $T_1$ and $T_2$. The union is performed on the adjacent cluster preserving the topology tree properties defined above.

When new edge is added, the two separated tree is combined to form a single tree. If the degree of the vertex after union exceed from 3 then the deeply nested cluster is then split till root to preserve the topology tree properties.

The figure next page defines the hierarchical topological partitioning and the analogous topology tree.

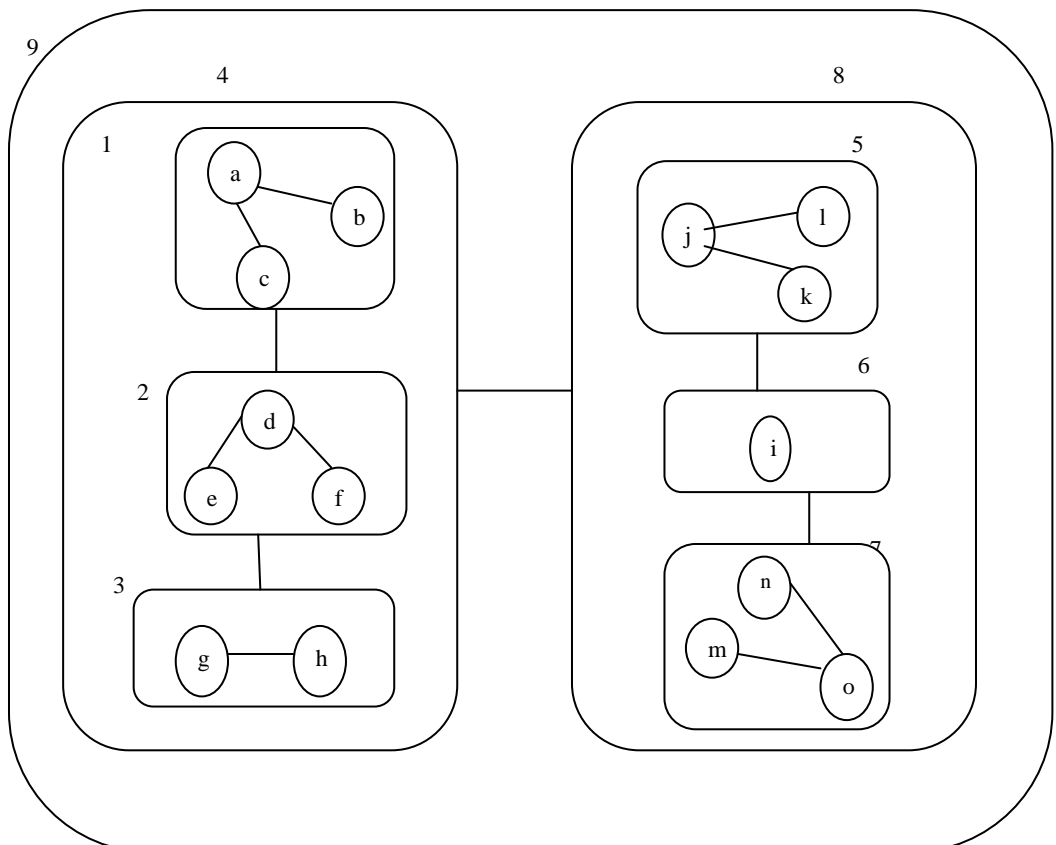

**Figure1: The hierarchical Topological partition**

- The tree define the node at level one as a cluster at level 1 which is the root node containing the single vertex.

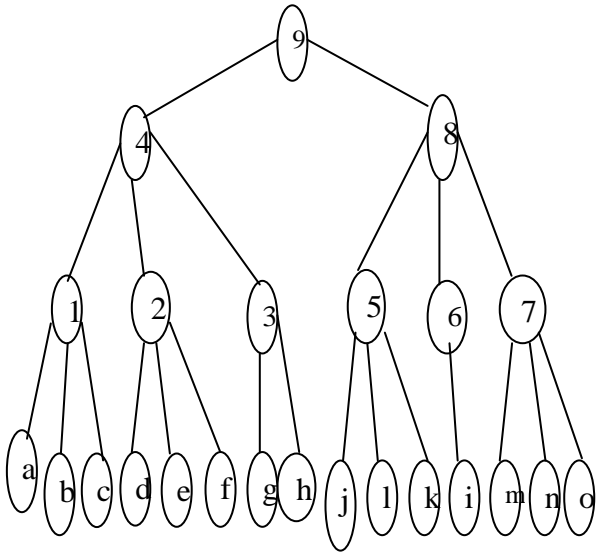

**Figure 2 Topology Tree.**

The height of the topology tree is O (logn). Hence any update in tree involves some level and require the local adjustments, so time required to update the topology tree is O (logn).

## 3.2 Euler Tour Tree Data Structure
The Euler tour is one of the data structure used to implements the randomized algorithmic techniques. The tree T is encrypted with the v nodes in the tree and any random node can be chosen as a root node.

The **Euler tour** visit the every edge exactly once and if represented as a tree then visiting every edge exactly twice, once entering into the vertex and once leaving that vertex.

**Definition**: The Euler tour tree reserve the Euler tour of the tree and represent the Euler tour in the balance binary search tree.

Euler tour tree are the substitution for the link-cut trees. These trees are apparent and easier to evaluate than the link cut trees for dynamic graphs.

The tree does not store the path information about the trees but store the procure information on the sub trees. The tour is the depth first traversal of the tree which return to the root node at the end.

The figure shows the Euler tour of the tree, directed edges show the sequence of visitation.

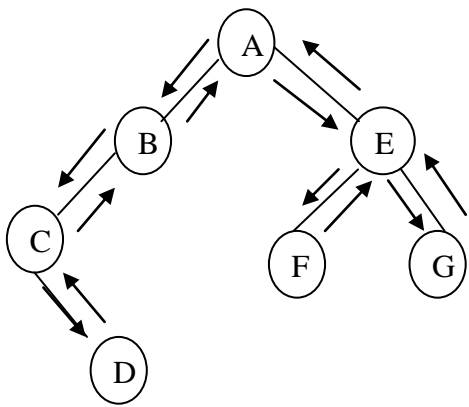

**Figure 3 Euler tour of tree**

The sequence of visitation here is  A , B , C ,D ,C,B,A ,E ,F ,E ,G ,E ,A for the tree T. The tour started and ended with the root node A.

The Euler tour function is called for the visitation where the function E(s) is defined as:

- E(r)
- Visit the vertex r.
- While every children x of r
-     Do call E(x)
- Visit r.

Here, the d-degree node is visited d times and edges twice except the root node which is visited d+1 times. The function E (T) represents the sequence of Tree T. Adjacency list with arrays and pointers are used to store the vertex occurrence.

The node holds the pointer to the visited sequence in the BST presenting the first and last time it was visited. The figure shows it below:

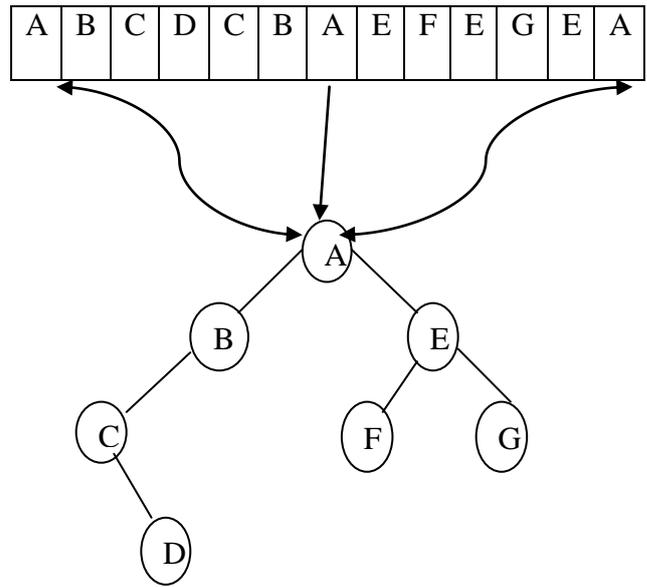

**Figure 4 pointer related to root.**

*3.2.1 Operations*
The Euler tour tree supports the following operations:
- **Cut(x):** cut the sub tree rooted at x while splitting the BST before the first visit and last visit to x and concatenating the both[10].
- **Find root(x):** returning the root of the node x, where root is visited first and last.
- **Link(x, y):** it insert x sub tree as a child of node y.

Every operation in Euler tour can preserves the property by splitting, merging and searching in the Euler tour tree. The operation involves the O (log n) per operation for the update in the Euler tour tree.

## 3.3 Dynamic Trees
There are number of dynamic trees which are used for dynamic graph algorithms to maintain the graph properties. One of them is **link-cut trees** which have the application in the area of Network problems and dynamic connectivity problem. These trees were introduced by Sleator and Tarjan [5]. The tree maintain the logarithmic amortized time per update operation. For details refer to [5].

## 3.4 Top Trees
The tree was introduced by alstrup [7].
The top tree support the path oriented updates and Queries basically for the problem of Divide and conquer algorithm. The working of top tree depends on boundary nodes and clusters.

Maintaining top trees of height $O(\log n)$ and with $O(m^{1/2})$ cluster nodes supporting Link, Cut, and Expose above with a sequence of $O(\log n)$ Merge and Split and $O(1)$ create and Destroy operations per update. The sequence itself is computed in $O(\log n)$ time.

The figure shows the case of two Clusters A and B and the parent node C in the top tree.

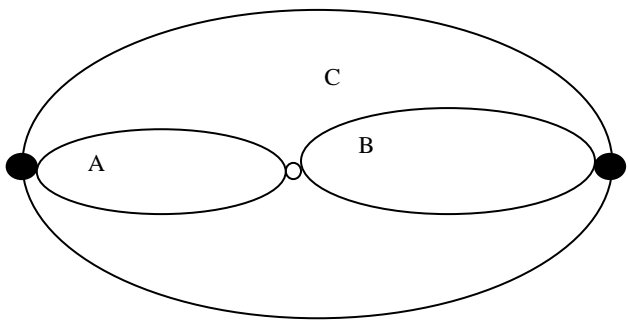

**Figure 5 cluster with boundary nodes**

## 4. COMPARITIVE ANALYSIS OF DYNAMIC GRAPH ALGORITHMIC TECHNIQUES

The following comparative analysis if performed on undirected graph G (m,n) is using various dynamic graph algorithmic techniques. Here k is the logarithmic of n.

**Table 1 Graph updates for techniques**

| Technique / Graph updates | Clustering | Sparsification | Randomization |
|---|---|---|---|
| Minimum spanning forest | $O(k^3)$* | $O(n^{1/2})$ | $O(\log 3n)$ |
| 2-edge connectivity | $O(mk^3)$ | $O(n^{1/2})$ | $O(\log 3n)$ |
| Bipartition Of graph | $O(m^{1/2})$ | $O(n^{1/2})$ | $O(\log 3n)$ |
| Space | $O(m \log n)$ | $O(m \log n)$ | $O(m+n \log n)$ |

**Table 2 Query time for techniques**

| Techniques / Query time | Clustering | Sparsification | Randomization |
|---|---|---|---|
| Minimum Spanning forest | $O(1)$ | $O(1)$ | $O(\log 3n)$ |
| 2-edge connectivity | $O(\log n)$ | $O(\log n)$ | $O(\log n/\log \log n)$ |
| Bipartition Query | $O(\log n)$ | $O(n^{1/2})$ | $O(1)$ |

Here, the randomized algorithm work well and polylogarithmic update time for various graph Properties and sparsification provides the better query time.

## 5. CONCLUSION
The paper studies the various algorithmic techniques for the undirected graph and provides the detail discussion and comparison among the techniques for retaining the various graph properties.
The data structure tools are defined in the paper is the underlying base for various graph techniques.

## 6. FUTURE SCOPE
In the process of algorithmic techniques and data structure, improvements can be done at various steps.
- Like solution can be provided for efficiently fully dynamic single-source reach ability and shortest paths on general graphs.
- These techniques can be combined together with other techniques to give better lower time bound and giving fully dynamic algorithm than increase only.
- Furthermore, no randomized algorithm is known for fully dynamic maintenance for shortest path. Future work can be done for finding out the randomized algorithm for faster solution.

On the practical side, it would be interesting to push these designs into real systems and real deployments of networking problems and attacks. The Sybil attack in distributed systems refers to individual malicious users joining the system multiple times under multiple fake identities. Sybil attacks can easily invalidate the overarching prerequisite of many fault-tolerant designs which assume that the fraction of malicious nodes is not too large.

Beyond Sybil attacks and beyond social networks, the insights on attack edges, cuts, and mixing time may find applications elsewhere using these techniques of undirected graph on network problem. For example, these insights might apply to Page Rank and help it to be robust against Sybil webpage's. One could also imagine detecting email spams based on the email graph and its connectivity property.

# 7. REFERNCES

[1] FREDERICKSON, G. N. 1985. Data structures for on-line updating of minimum spanning trees. *SIAM J,* 781–798.

[2] EPPSTEIN, D., GALIL, Z., ITALIANO, G. F., AND NISSENZWEIG, A. 1997. Sparsification–A technique for speeding up dynamic graph algorithms. *J. ACM 44*, 5 (Sept.), 669 – 696.

[3] Monika R. Henzinger and Valerie King. Randomized fully dynamic graph algorithms with polylogarithmic time per operation. J. ACM 46(4) (1999), 502–516.

[4] GALIL, Z., AND ITALIANO, G. F. 1992. Fully dynamic algorithms for 2-edge connectivity. *SIA J. Comput. 21*, 1047–1069.

[5] D. D. Sleator, R. E. Tarjan, A Data Structure for DynamicTrees, Journal. Comput. Syst. Sci., 28(3):362-391, 1983.

[6] HENZINGER, M. R. 1995. Fully dynamic biconnectivity in graphs. *Algorithmica 13*, 503–538.

[7] Stephen Alstrup, Jacob Holm, Kristian De Lichtenberg,and Mikkel Thorup, *Maintaining information in fully dynamic trees with top trees*, ACM Transactions on Algorithms (TALG)

[8] Sauta Elisa Schaeffer, "Survey Graph clustering," ElsevierComputer Science Review, vol. I, pp. 27-64, 2007

[9] Reena Mishra, Shashwat Shukla, Dr. Deepak Arora and Mohit Kumar. Article: An Effective Comparison of Graph Clustering Algorithms via Random Graphs. *International Journal of Computer Applications* 22(1):22–27, May 2011. Published by Foundation of Computer Science.

[10] G. W. Flake, R. E. Tarjan, and K. Tsioutsiouliklis. Graph clustering and minimum cut trees, Internet Mathematics, 1(3), 355-378, 2004.

[11] Mikkel Thorup: Near-optimal fully-dynamic graph connectivity. STOC 2000: 343-350

[12] Jacob Holm, Kristian de Lichtenberg, Mikkel Thorup: Poly-logarithmic deterministic fullydynamic algorithms for connectivity, minimum spanning tree, 2-edge, and biconnectivity. J. ACM 48(4): 723-760 (2001).

[13] Mihai Patrascu, Erik D. Demaine: Lower bounds for dynamic connectivity. STOC 2004: 546-553.

[14] G. N. Frederickson. A data structure for dynamically maintaining rooted trees. Journal Of Algorithms, 24(1):37{65, 1997.

[15] U. A. Acar, G. E. Blelloch, and J. L. Vittes. An experimental analysis of change propagation in dynamic trees. pages 41{54, 2005.

[16] R. C. K. and S. Sudarshan. Graph clustering for keyword search. In COMAD ' 09, 2009.